\documentclass{aa}
\usepackage{amsmath,graphicx,amssymb}
\usepackage{txfonts}
\usepackage{amssymb}
\usepackage{natbib}
\bibpunct{(}{)}{;}{a}{}{,}
\def\urltilda{\kern -.15em\lower .7ex\hbox{\~{}}\kern .04em}

\newcommand{\hzav}[1]{\left[#1\right]}

\newlength\staretab

\makeatletter
\def\sgn{\mathop{\operator@font sgn}\nolimits}
\makeatother

\begin{document}

\title{The (non-)variability of magnetic chemically peculiar candidates in the
    Large Magellanic Cloud}
\titlerunning{mCP candidate stars in the LMC}

\author{E. Paunzen\inst{1,2}
    \and Z.~Mikul\'a\v sek\inst{1,3}
    \and R.~Poleski\inst{4,5}
    \and J.~Krti\v{c}ka\inst{1}
    \and M.~Netopil\inst{6}
    \and M.~Zejda\inst{1}}
\authorrunning{E.~Paunzen et al.}
\offprints{E. Paunzen,\\
\email{epaunzen@physics.muni.cz}}

\institute{Department of Theoretical Physics and Astrophysics,
           Masaryk University, Kotl\'a\v{r}sk\'a 2, CZ\,611\,37, Brno,
           Czech Republic
        \and Rozhen National Astronomical Observatory, Institute of Astronomy of
        the Bulgarian Academy of Sciences, P.O. Box 136, BG-4700 Smolyan, Bulgaria
        \and Observatory and Planetarium of Johann Palisa, V\v SB --
            Technical University, Ostrava, Czech Republic
        \and Warsaw University Observatory, Al. Ujazdowskie 4,
        00-478 Warszawa, Poland
        \and Department of Astronomy, Ohio State University, 140 W. 18th Ave.,
        Columbus, OH 43210, USA
        \and Institut f{\"u}r Astrophysik der Universit{\"a}t Wien,
           T{\"u}rkenschanzstr. 17, A-1180 Wien, Austria}

\date{Received 2012/ Accepted 2013}

\abstract{The galactic magnetic chemically peculiar (mCP) stars of the upper main sequence are well known as periodic spectral and light variables. The observed variability is obviously caused by the uneven distribution of overabundant chemical elements on the surfaces of rigidly rotating stars. The mechanism causing the clustering of some chemical elements into disparate structures on mCP stars has not been fully understood up to now. The observations of light changes of mCP candidates recently revealed in the nearby Large Magellanic Cloud (LMC) should provide us with information
about their rotational periods and about the distribution of optically active elements on mCP
stars born in other galaxies.}
{We queried for photometry at the Optical Gravitational Lensing Experiment (OGLE)-III survey
of published mCP candidates selected because of the presence of the characteristic $\lambda5200\AA$
flux depression. In total, the
intersection of both sources resulted in twelve stars. For these objects and two control
stars, we searched for a periodic variability.}
{We performed our own and standard periodogram time series analyses of all available data. The final
results are, amongst others, the frequency of the maximum peak and the bootstrap
probability of its reality.}
{We detected that only two mCP candidates, 190.1 1581 and 190.1 15527, may show some weak
rotationally modulated light variations with periods of 1.23 and 0.49 days; however, the 49\% and 32\% probabilities
of their reality  are not very satisfying. The variability of the other 10 mCP candidates is too low to be detectable
by their $V$ and $I$ OGLE photometry.}
{The relatively low amplitude variability of the studied LMC mCP candidates sample can
be explained by the absence of photometric spots of overabundant optically active
chemical elements. The unexpected LMC mCPs behaviour is probably caused by different conditions
during the star formation in the LMC and the Galaxy.}


\maketitle

\section{Introduction}

The chemically peculiar (CP) stars of the upper main sequence display
abundances that deviate significantly from the standard abundance
distribution. For a subset of this class, the magnetic chemically peculiar
(CP2 or mCP) stars, the
existence of strong global stellar magnetic fields was found.

The variability of mCP stars is explained in terms of the oblique
rotator model \citep{sti50}, according to which the period of the
observed light, spectrum, and magnetic field variations is the
rotational period. The photometric changes are due to
variations of global flux redistribution caused by the phase-dependent
line blanketing and continuum opacity namely in ultraviolet part of
stellar spectra \citep{krt901,krtcuvir}.

The amplitude of the photometric variability is a combination of the
characteristics of the degree of nonuniformity of the surface
brightness (spots), the used pass band, and the line of sight.
The observed amplitudes are up to a few tenths of magnitudes. However,
for some stars
one also fails to find any rotational induced variability at all.
The locations of spots used to have connection with
the dipole-like magnetic field geometry.

In the Milky Way, we know of a statistically significant number of
rotational periods for mCP stars deduced from photometric and/or
spectroscopic variability studies \citep{reca01,zoo}.
Nevertheless, extragalactic mCP stars were also found.
After the first photometric detection of classical CP
stars in the Large Magellanic Cloud \citep{mai01}, a long-term effort
was made to increase the sample \citep{pau06}. We were, finally, able
to verify our findings with spectroscopic observations \citep{sipau}.

In this paper, we present the time series analysis of light
variations of photometrically detected mCP candidates in the LMC. Our
list of targets \citep{pau06} was compared with the OGLE database
\citep{udal} for corresponding measurements. In total, fourteen
common objects were found and their $V$ and $I$ light curves
were analysed.

\begin{table*}
\begin{center}
\caption{The basic photometric parameters from \citet{pau06} and references therein, as
well as the photometric characteristics of CP candidates in the $V$ and $I$ bands.
$\overline{V}$ is the mean magnitude in $V$ filter, $\overline{M}_{V0}$ the $V$ absolute
magnitude assuming the distance modulus $m-M=18.49\pm0.05$\,mag \citep{piet} and $A_V=0.25$\,mag \citep{sipau},
$(\overline{V-I}) $ the mean colour index, $\Delta a$ the mean Maitzen index,
$N_{\rm{V}}$ and $N_{\rm{I}}$ the numbers of $V$ and $I$ observations, $\delta m_V$
and $\delta m_I$ the typical uncertainties of particular measurements, $1/P$ the
frequency of the maximum effective
amplitude of $A_{\rm{m}}$ or $S/N$, while $A_{\rm{m}}$ and $A_{\rm{ms}}$ are the observed
maximum modified amplitude and expected maximum modified amplitude. $S/N$ and $S/N_s$ are
the maximum signal/noise value and its expected value; ``prob'' means the probability
of the reality of designate frequency of periodic variations. Bold numbers designate higher values than the particular mean
values.} \label{hvezdy}\tiny
\begin{tabular}{rlcccccccccccccc}
\hline
 No.&  OGLE & $\overline{V}$ & $\overline{M_{V0}}$ & $(\overline{V-I}) $ & $\Delta a$ & $N_{V}$ &
 $\delta m_{V}$ & $N_{I}$ & $\delta m_{I}$ & $1/P$ & $A_{\rm{m}}$ & $A_{\rm{ms}}$ & $S/N  $ & \ $S/N_s$ & prob  \\
 &LMC + & [mag] & [mag]& [mag] & [mag] & & [mag] & & [mag] & [d$^{-1}$] & [mag] & [mag] & & & \% \\
\hline
 1&  135.3 4273* & 17.80 & $-$0.95 &  +0.85 & (0.089) & 67 & 0.013 & 370 & 0.013 & 1.1336 & 0.175  & 0.041 & 51  & 4.8 & 100\\
 2&  135.3 30107*& 17.76 & $-$0.98 &  +1.36 & (0.094) & 65 & 0.011 & 466 & 0.010 & 0.6522 & 0.005 & 0.006 & 4.5 & 4.5 & 16\\
 3&  136.7 861   & 17.60 & $-$1.14 &  $-$0.04& 0.078 & 28 & 0.010 & 301 & 0.018 & 1.0088 & 0.014 & 0.013 & 5.1 & 4.5 & 29\\
 4&  136.7 16501 & 19.16 & +0.41  &  +0.03 & 0.087 & 24 & 0.024 & 266 & 0.058 & 0.2286 & 0.044 & 0.040 & 5.1 & 4.6 & 28\\
 5&  136.8 678   & 17.84 & $-$0.90 &  +0.01 & 0.095 & 45 & 0.011 & 435 & 0.021 & 1.4388 & 0.013 & 0.013 & 4.7 & 4.6 & 18\\
 6&  136.8 1801  & 18.79 & +0.05  &  +0.11 & 0.085 & 45 & 0.019 & 437 & 0.041 & 1.1811 & 0.027 & 0.023 & 5.3 & 4.3 & 30\\
 7&  136.8 1873  & 18.64 & $-$0.10  &  +0.02 & 0.056 & 45 & 0.017 & 437 & 0.039 & 0.6549 & 0.021 & 0.022 & 4.2 & 4.3 & 20\\
 8&  136.8 2002  & 18.67 & $-$0.08  &  +0.10 & 0.054 & 45 & 0.018 & 434 & 0.038 & 0.4372 & 0.020 & 0.022 & 4.1 & 4.5 & 13\\
 9&  136.8 3694  & 19.10 & +0.35  &  +0.05 & 0.052 & 44 & 0.023 & 421 & 0.056 & 0.9838 & 0.031 & 0.033 & 4.4 & 4.7 & 23\\
10&  136.8 3875  & 18.98 & +0.24  &  +0.04 & 0.087 & 43 & 0.022 & 434 & 0.052 & 1.8641 & 0.031 & 0.032 & 4.6 & 4.7 & 10\\
11&  190.1 1445  & 18.08 & $-$0.65 &  $-$0.03& 0.095 & 52 & 0.013 & 454 & 0.037 & 0.9183 & 0.017 & 0.016 & 5.6 & 5.3 & 26\\
12&  190.1 1581  & 17.89 & $-$0.85 &  +0.01 & 0.044 & 51 & 0.012 & 454 & 0.031 & 0.8075 & 0.019 & 0.015 & 7.1 & 5.3 & 49\\
13&  190.1 2822  & 18.68 & $-$0.06  &  +0.02 & 0.060 & 53 & 0.018 & 454 & 0.056 & 1.6884 & 0.022 & 0.024 & 4.6 & 5.1 & 11\\
14&  190.1 15527 & 17.16 & $-$1.57 &  +0.01 & 0.041 & 53 & 0.008 & 453 & 0.019 & 2.0260 & 0.011 & 0.009 & 6.7 & 5.5 & 32\\
  \hline
\end{tabular}
\end{center}
*The control stars: the first is a short-periodic Cepheid, the
second a non-variable K giant.
\end{table*}

\section{Light variations of LMC mCPs candidates}

Overabundant chemical elements are very unevenly distributed
on the surfaces of
Galactic mCP stars which results in the periodic variations of their
spectra and brightness. The goal
of the following analysis is to find rotationally modulated light
changes and to ascertain their rotational periods. The basic data
representing 6476 individual photometric measurements were taken from the
Optical Gravitational Lensing Experiment (OGLE)-III survey of the LMC
\citep{udal}, with 90\% taken in the $I$ band and the remaining 10\%
in the $V$ band (see Table \ref{hvezdy}).

\subsection{Target selection and description of the OGLE LMC data}

The LMC mCPs candidates were selected on the basis of $\Delta a$
photometry. Thanks to the typical flux depression in CP stars at
$\lambda$~5200\,\AA, the tool of $\Delta a$ photometry is able to
detect them economically and very efficiently by comparing the
flux at the centre (5200\,\AA, $g_2$) with the adjacent regions
(5000\,\AA, $g_1$ and 5500\,\AA, $y$). It was shown that virtually all
chemically peculiar stars with magnetic fields have significant
positive $\Delta a$ values up to +100\,mmag whereas Herbig Be/Ae and
metal-weak stars exhibit significantly negative values \citep{Pa2005b}.

We compared our list of mCP candidate stars \citep{pau06} with the OGLE database
for corresponding measurements on the basis of equatorial coordinates
and $V$ magnitudes. After a first query, we inspected the positions in
our original images and those of OGLE by eye. In total, we found
fourteen matches in both sources. The final list of stars is given in
Table \ref{hvezdy}.

\begin{figure}[t]
\centering \resizebox{0.837\hsize}{!}{\includegraphics{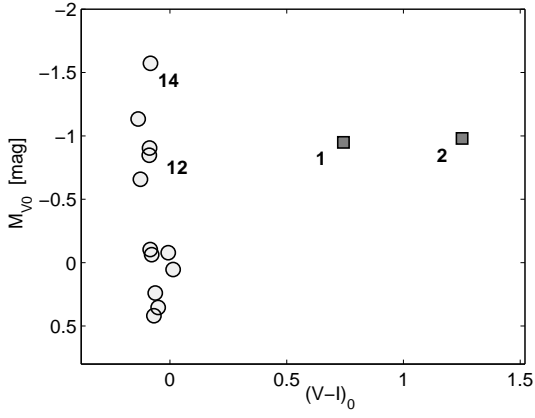}}
\caption{The Hertzsprung-Russell diagram of all analysed stars. $M_{V0}$ is a mean $V$ absolute magnitude corrected for extinction $A_V=0.25$\,mag \citep{sipau} assuming the distance modulus $m-M=18.49$\,mag as given by \citet{piet}; $(V-I)_0$ is the colour index corrected for the corresponding interstellar reddening by 0.1\,mag. Open circles clustering towards the main sequence indicate mCP candidates, the square denoted by 1 is the mean position of star 1 (135.3 4273), which is a short-period Cepheid. Star 2 (135.30107) is a normal non-variable K giant. Stars 12 and 14 are mCP candidates suspected for their rotationally modulated light variations.} \label{HR}
\end{figure}

A further analysis revealed that the sample of mCP
candidates is contaminated by two late-type stars. The
first, 135.3 4273, was recognised as the short-period Cepheid
OGLE-LMC-CEP-0327 pulsating in its first overtone with the period
$P=0.8821659(13)$~d \citep{sos08}. The second, 135.3 30107, is a
normal, non-variable K-type giant. We adopted them as control stars
and their photometric data were analysed together with other stars of
the sample. The Hertzsprung-Russell diagram of all studied stars constructed from data
given in Table \ref{hvezdy} is shown in Fig. \ref{HR}.

The data of the mCP candidates contain the JD$_{\mathrm{hel}}$ date of
measurement, $t_i$, the measured $V$ and $I$ magnitudes $m_i$, 
and the estimate of its internal uncertainty $\delta m_i$. We have found that
the last quantity represents the real uncertainty of the magnitude
determination quite well and that is why we used it to weight the
individual measurements according to the relation
$w_i\sim\delta m_i^{-2}$.

\subsection{mCP star candidates}

The mean $V$ magnitudes of mCP candidates (stars 3--14)
span the interval from 17.16 to 19.16 mag. Assuming a distance
module ($m-M$) for the LMC of 18.33 mag and an extinction of
$A_{\rm V}=0.25$~mag \citep{sipau}, we obtain the interval of absolute
$V$ magnitudes from $-$1.4 to +0.6~mag, respectively. It corresponds
mainly to Si-type of mCP stars, spectroscopically verified by
\citet{sipau}.

The median of the amplitude of light variations of mCP stars in $V$ is
about 0.032 mag \citep{zoo}. The typical error of the determination of
one $I$ measurement of our mCP candidate set is $\sigma_{\rm
It}\sim 0.038$~mag and the typical number of measurements is $N_{\rm
It}\sim 440$ (see Table \ref{hvezdy}). Consequently, the expected
uncertainty of $\delta_{\rm It}$ of the amplitude determination of
periodic variations can be estimated as $\delta_{\rm
It}=\sqrt{8/N_{\rm It}}\,\sigma_{\rm It}\sim 0.005$~mag. The same
quantities for $V$ measurements are $\sigma_{\rm Vt}\sim 0.015$~mag and
$N_{\rm Vt}\sim 45$, $\delta_{\rm Vt}=\sqrt{8/N_{\rm Vt}}\,\sigma_{\rm
Vt}\sim 0.006$~mag. We found this encouraging for the project of detecting
periodicities in OGLE-III photometry.

Before starting the analysis of $V$ and $I$ photometric data we investigated the
expected relationship between infrared and visual variations of mCP stars in more detail.

\subsection{$I$ light variability of mCP stars}\label{teorie}

Our search for periodic rotationally modulated variations in the sample of 12 LMC mCP
candidates is based mainly on the analysis of the variability of these stars in the near infrared,
while the Galactic mCP star variability is studied mainly in the visual region. Unfortunately, the
information about the infrared variability of Galactic mCP stars is very scarce. From the
few observations available \citep[e.g.][]{musi,catlek,wraight} one can conclude, that the amplitude
of the light variations in the near infrared filters is similar but not identical to that in visual filters. This can be
easily understood from the theoretical model of the light variability of mCP stars \citep{mol75,krt901,seuma},
which is based on the light redistribution from the shorter wavelength region (typically the far ultraviolet)
to the longer wavelength region (typically the visual and infrared) because of enhanced opacity in the regions
with overabundant elements. The infrared continuum lies in the Rayleigh-Jeans part of the flux distribution
function, where the ratio of the two different fluxes is independent of wavelength.

To derive a quantitative prediction for the mCP star light variability
in the near infrared, we employed successful models of the visual
light variability of \object{HD 37776}
\citep[$T_\text{eff}=22\,000\,\text{K}$]{krt901}, \object{HR 7224}
\citep[$T_\text{eff}=14\,500\,\text{K}$]{krt7224}, and \object{CU Vir}
\citep[$T_\text{eff}=13\,000\,\text{K}$]{krtcuvir} and predicted the
light curves in the $I$ filter. The light curves are calculated from
the surface abundance maps of these stars \citep{kus,choch,leh2} using
TLUSTY model atmospheres and SYNSPEC spectrum synthesis code
\citep{lahub,bstar2006}. For these stars, no observations in $I$
are available.

The light curves are predicted using specific intensities filtered by
appropriate transmission curve. We fitted the OGLE transmission curve
by a suitable formula with a precision better than 3\% in the form of
\begin{equation}
M(\lambda)=\left\{
 \begin{array}{cl}
  \exp\,(a_1x+a_2x^2), & \lambda<\lambda_0,\\
  \exp\,(a_4x+a_5x^2+a_6x^3), & \lambda>\lambda_0,
 \end{array}\right.
\end{equation}
where the variable $x$ is connected with the wavelength $\lambda$ in
\AA\ as
\begin{equation}
x=\left(\frac{\lambda-\lambda_0}\sigma\right)^2,
\end{equation}
and
\begin{subequations}
\begin{align}
\lambda_0&=8111\,\AA,&\sigma&=733\,\AA,\\
a_1&=-0.501,&a_2&=-0.329,\\
a_4&=-0.400,&a_5&=+0.452,&a_6&=-1.03.
\end{align}
\end{subequations}

\begin{figure}[t]
\centering \resizebox{0.66\hsize}{!}{\includegraphics{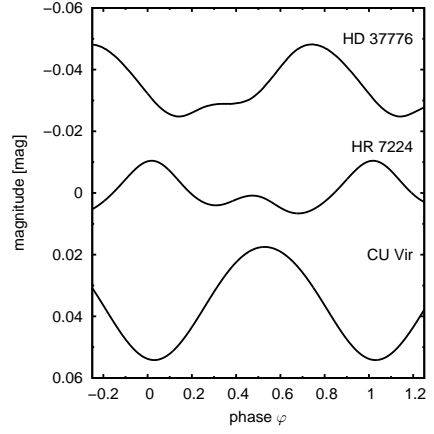}}
\caption{The predicted light curves in the $I$ band of selected Galactic
mCP stars.} \label{hvvelir}
\end{figure}

The resulting light curves are given in Fig.~\ref{hvvelir}. From this
plot we conclude that if the LMC mCP stars have the same
amplitude as their Galactic counterparts, their light variations
should be detectable by OGLE.

\section{Search for photometric periodicity}

The detection of photometric variations of LMC mCP candidates is not straightforward because
of their faintness. That is why we approximated their light behaviour by
the simplest possible model assuming that light curves in all filters are simple sinusoidal with
the same amplitude. Whenever a signal of this kind is detected, we are able to improve the model even more.

\subsection{Periodograms}

The periods of variable stars are usually inferred by means of various
kinds of periodograms. We opted for two versions of periodograms based on the fitting of detrended
measurements $y_i$ with uncertainties $\sigma_i$ done in the moments $t_i$ by an ordinary sine-cosine model,
$f(\omega,t)= b_1\,\cos(\omega\,t)+b_2\,\sin(\omega\,t)$, where $ \omega=2\pi/P $ is an angular frequency and
$P$ the period, using the standard $\chi^2$ least-squares method. If the sum $\chi^2(\omega)$ is for the given
$\omega$ minimum the so-called \emph{modified amplitude} $A_{\mathrm{m}}(\omega)$ has to be maximum,
\begin{eqnarray}\label{Am}
&\displaystyle \sum^N_{i=1} \frac{y_i^2}{\sigma_i^2}-\chi^2(\omega) =\sum^N_{i=1}\hzav{\frac{f(\omega,t_i)}
{\sigma_i}}^2= A^2_{\mathrm{m}}\sum^N_{i=1} \frac{1}{8\,\sigma_i^2},\quad \Rightarrow\\
&\displaystyle A_{\mathrm{m}}(\omega)=\sqrt{\frac{8}{\sum \sigma_j^{-2}}\sum^N_{i=1}
\hzav{\frac{b_1(\omega)\cos(\omega t_i)+b_2(\omega)\sin(\omega t_i)} {\sigma_i}}^2},
\end{eqnarray}
where $b_1(\omega)$ and $b_2(\omega)$ are coefficients of the fit. When there is a uniform phase
coverage, the modified amplitude is equal to the amplitude of sinusoidal signal.
The best phase sorting of the observed light variations corresponds to the
angular frequency $\omega_{\rm{m}}$ with the maximum of modified amplitude $A_{\rm m}(\omega)$.

The second LSM type of periodogram uses for the significance of individual peaks a
robust signal-to-noise ($S/N$) criterion, which is defined as
\begin{equation}\label{S/N}
S/N(\omega)=\frac{Q(\omega)}{\delta Q(\omega)}=\frac{b_1^2(\omega)+
b_2^2(\omega)}{\delta\hzav{b_1^2(\omega)+b_2^2(\omega)}},
\end{equation}
where $\delta Q(\omega)$ is an estimate of the uncertainty of the quantity $Q(\omega)$ for a particular angular frequency.

We tested the properties of the $S/N(\omega)$ criterion a thousand samples with sine signals scattered by randomly
distributed noise. We found that if there is no periodic signal in this data, the median of the maximum $S/N(\omega)$ value
in a periodogram is 4.52; in 95\% of cases we find a $S/N$ value between 4.2 and 5.4. The occurrence of peaks definitely higher
than 6 indicates possible periodic variations. The detailed description of both LSM novel periodogram criteria will be published elsewhere.

During the treatment of OGLE-III time series, we concluded that both types of periodograms correlate very well with other
time-proven estimates, the Lomb-Scargle \citep[see e.g.][]{rybicki} periodogram, for example. So we are able to consider them as generally interchangeable.

We tested all frequencies from 0 to 2.1\,d$^{-1}$. The upper
limit is slightly above the frequency of the fastest rotating CP star known to date \citep[HD\,164429, with the period of $P=0\fd51889$, see][]{adel}.

\begin{figure}[t]
\centering
\resizebox{1.00\hsize}{!}{\includegraphics{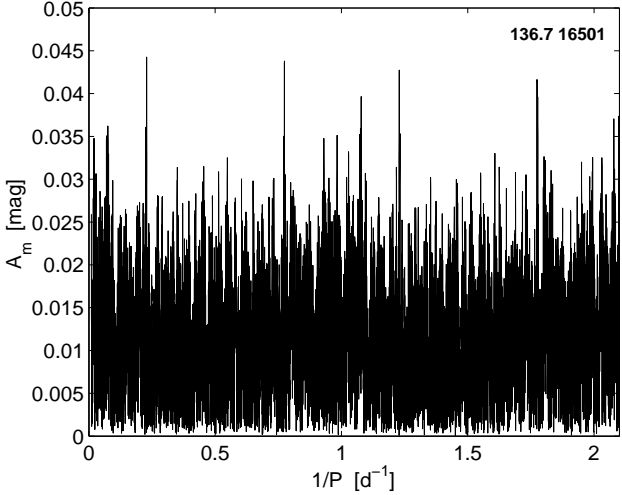}}
\caption{A typical periodogram of a mCP candidate, star 4
(136.7.16501); all peaks are inconspicuous.}
\label{per_1}
\end{figure}

\begin{figure}[t]
\centering \resizebox{1.\hsize}{!}{\includegraphics{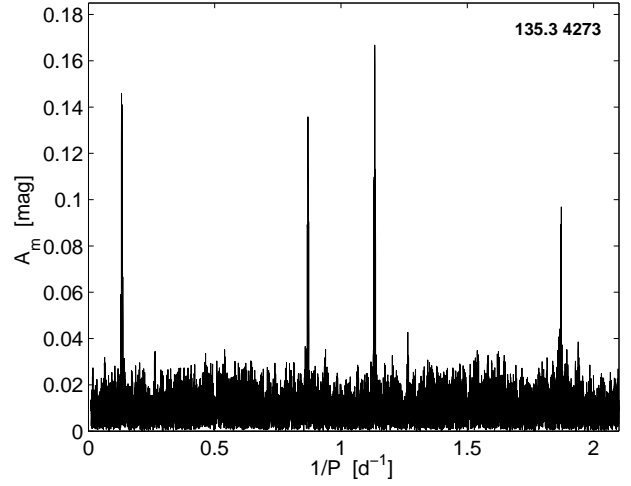}}
\caption{The basic period of the overtone-Cepheid OGLE-LMC-CEP-0327 (135.3\ 4273) plus
its aliases conjugated with the period of a sidereal day.}
\label{per_cef}
\end{figure}

\subsection{Discussion of periodograms}

We constructed periodograms for all stars of our sample and calculated frequencies
$f=1/P$ of their maximum peaks and modified amplitudes $A_{\rm m}$ and $S/N$s for these peaks.
These quantities are given in Table \ref{hvezdy}. We compared
the periodograms of the LMC mCP candidates with those of the control stars, where we clearly detected
the variability of OGLE-LMC-CEP-0327 and its aliases conjugated with the period of a sidereal
day (Fig.\,\ref{per_cef}). Our results exactly agree with the previous determination of the
period and other characteristics of the star. This lends confidence to our time series analysis method.

\begin{figure}[t]
\centering \resizebox{0.968\hsize}{!}{\includegraphics{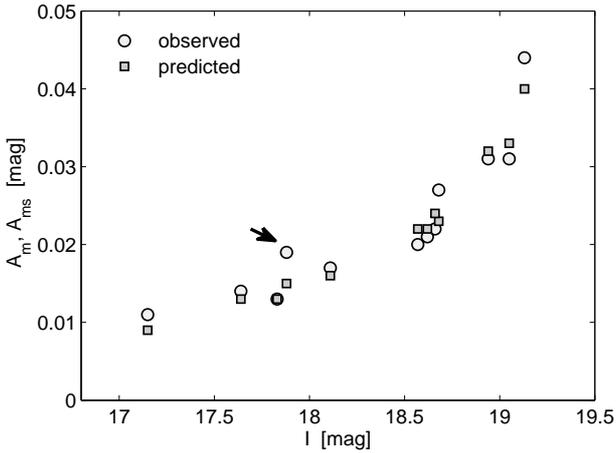}}
\caption{The dependence of the maximum modified amplitudes of the observed ($A_\mathrm{m}$ -- circles)
and the shuffled data ($A_\mathrm{ms}$ -- squares) on the mean $I$ magnitude of mCP candidates. The
largest positive relative deviation is found for the star 12 (190.1\,1581) and is denoted by an arrow.} \label{obs_shu}
\end{figure}

The periodograms of the known Cepheid and the mCP candidates plus the second control star
apparently differ: the peaks here have only a very narrow margin to the body of a pure scatter
(compare Fig.\,\ref{per_1} and Fig.\,\ref{per_cef}). There are additional arguments to explain why
all periods except those found for the stars 12 and probably star 14, listed in Table \ref{hvezdy},
are very likely only coincidences:

\begin{itemize}
\item The distribution of the maximum amplitude frequencies is shallow
    and the found frequencies cover the studied interval 0 to 2.1
    d$^{-1}$ more or less evenly. This is a sign that observed peaks are
    most likely mere coincidences.
\item The median of periods of maximum amplitude, 1.0~d, differs
    significantly from the median period of Galactic mCP stars, 3.2~d \citep{zoo}.
\item The amplitudes of light changes of Galactic mCP stars display no correlation with
colour index, spectral type, or absolute magnitude. It is in sharp contrast with observations of LMC mCP candidates whose
    observed maximum amplitudes exhibit strict monotonic dependence on the mean magnitude
    (see circles in Fig.\,\ref{obs_shu}). It suggests that the nature of observed
    variability of LMC objects is different.
\item It is conspicuous that the median of $S/N$ ratios of the highest peaks in periodograms of individual CP
candidates is only 4.7, corresponding to the median of maxima peaks $S/N=4.52$ in the case of pure scatter. It indicates
the lack of detectable periodic signal in the majority of inspected mCP candidates, with two exceptions (stars 12 and 14)
which are discussed in the following.
\end{itemize}

This persuaded us to test the hypothesis that the distribution of the data is random
via a heuristic \emph{shuffling method}. We re-analysed the data of each mCP candidate in the
same way as described above, only we randomly shuffled all the individual observations
(magnitudes and their uncertainties), the times of the observations remained the same. We are
convinced that any periodic signal had to be destroyed. We calculated $A_{\rm{ms}}$ and $S/N_{\rm{s}}$
for each shuffled version
of data. The median of the both shuffled quantities is
given in Table \ref{hvezdy} and Fig.\,\ref{obs_shu}. It is apparent that the results 
are nearly the same as in the case of the original, unshuffled data with
two exceptions (stars 12 and No 14).

As an added and independent test, we applied two different time series
analysis methods, namely the modified Lafler-Kinman method \citep{Hensb77}
and the Phase-Dispersion-Method \citep{Stell78}. All computations were done
within the programme package Peranso\footnote{http://www.peranso.com/}.
Using these utilities we find only upper limits for variability, but no statistically significant
detection except for the case of star 1. The frequencies discussed in the following
for stars 12 and 14, are detected on a 3.4 and 3.9 $\sigma$ level, respectively.

\subsection{Significance of the found periods}

The characteristics of periodograms is strongly affected by the scatter of data; this forced us to
treat the time series very carefully and to develop special techniques that enabled us to extract hidden
information as effectively as possible.

\subsubsection{Aliasing}

The OGLE-III data used for our 12 LMC mCP candidates cover the whole observed time interval
rather unevenly and caused a lot of prominent aliases which make the spectrum of
periods relatively complex. Besides the dominant peak at the real frequency of variations, one may expect its
aliases to be conjugated with the frequency of the Earth's rotation and revolution around the Sun. However, the
schedule of OGLE exposures was more complicated.

Time series of the stars in our sample representing both 660 $V$ and
5816 $I$ measurements were obtained from 2001 to 2008. The complete photometry was done only when the
LMC was sufficiently high above the horizon. Consequently, 85\% of the measurements were obtained in the time
between 2.5 hours before and after the passage of the LMC through the local meridian. Similarly, 85\% of the
observations were done during six months in a year starting from mid-September until 
mid-March. There
is only a weak correlation between the number of measurements and the corresponding lunar phase. All
of this influenced the appearance of the details of the periodograms of the light variations of the individual stars.

\begin{figure}[t]
\centering \resizebox{1.05\hsize}{!}{\includegraphics{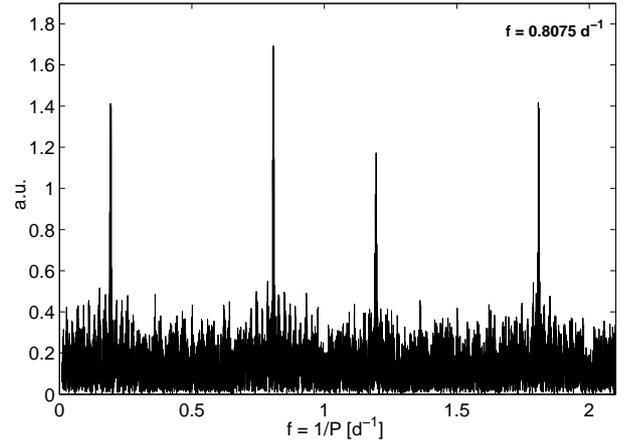}}
\caption{The simulated periodogram for star 12 (190.1\,1581) assuming a pure sine signal of the frequency
$f=1/P=0.8075\,\mathrm d^{-1}$. The most prominent peak corresponds to the basic frequency; the other pronounced peaks are
conjugated aliases with one sidereal day.} \label{alias}
\end{figure}

Because the determination of the position and significance is demanding, we used the method of the direct
modelling of aliases spectra by simulating the result of observations of the object with a given frequency of a sine signal.
The simulated periodogram of this situation then demonstrates the aliases spectrum and the mutual proportion of their
individual peaks. An example of a simulated spectrum of this kind for the basic frequency of $f=1/P=0.8075\,\mathrm d^{-1}$
is displayed in Fig. \ref{alias}. By default, the principal aliases are, double peaks, while the basic peak is a triplet
with a dominant central peak. This structure helped us to distinguish among real and alias peaks in the frequency spectrum.

We conclude that each star has its specific aliases spectrum with different proportions in the heights of individual
peaks and their inner structure.

\subsubsection{Bootstrap tests}

The bootstrap technique \citep{hall92} has proved to be very useful for testing the statistical significance and, therefore, the reality of found periods. It helped us to quantify this reality as a probability that the periodogram of a randomly
created bootstrap subset of the original
data has its dominant peak at the same frequency as the standard periodogram. We tested it with one hundred 
bootstrap subsets for each star of our sample. We consider a period to be statistically significant if the maximum peak occurs at one
of the aliased frequencies because during the bootstrap choice aliases often exceed the basic peak. The probability of
\emph{the period reality} for each star is given in the column ``prob'' in Table \ref{hvezdy}.

The results are rather dismal. With the exception of the undoubtedly variable star 1, the reality of finding a star period 
never exceeds 50\%. Nevertheless, two stars, 190.1 1581 and 190.1 15527, at least approach that limit.

\section{The mCP candidates suspected of periodic light variations}

Both objects suspected of periodic light variations, star 12 (190.1 1581) and star 14 (190.1 15527), display relatively low amplitude
light variation which are only slightly above the detectability by OGLE-III $V$ and $I$ photometry. Therefore, we used a very simple model
of their behaviour assuming the linear ephemeris and the sine form of the $V$ and $I$ variations with different amplitudes. We obtained the
following model of light changes
\begin{equation}\label{var}
\vartheta(t)=\frac{t-M_0}{P}, \quad m_c(\vartheta)=\overline{m}_c -\frac{a_c}{2} \cos(2\,\pi\,\vartheta),
\end{equation}
where $\vartheta(t)$ is a phase function, $M_0$ is the moment of the basic maximum, $P$ is the period in days, $m_c(t)$ is a
predicted magnitude in filter $c\ (c=V,\,I)$ at the time $t$, $\overline{m}_c$ is a mean value
of the magnitude in the filter $c$, and $a_c$ is an amplitude in the filter $c$. The values for both stars derived by standard least-squares minimization regression are given in Table \ref{hodnoty}.

\subsection{LMC 190.1 1581}

Star 12 (190.1 1581) is one of the brighter stars in our sample. Its mean absolute magnitude $\overline{M}_{V0}=-0.85$~mag and negative dereddened index $(\overline{V-I})_0=-0.09$~mag indicates that this object belongs among the late B-type stars. If it is a true mCP star, it is most
likely a Si- or He-weak type object.

In its periodogram we find a dominant peak at the frequency $f=0.8075~\rm d^{-1}$ corresponding to the period $P=1\fd 2433$.
The second peak of nearly the same height consists of two peaks at the frequencies $f=1.8075~\rm d^{-1}$ and
$f=1.81025~\rm d^{-1}$. Their distance of 0.00275~d$^{-1}$ corresponds to the reciprocal value of the sidereal
year in days. The internal structure indicates that it is an alias of the basic period. The bootstrap test of the
reality of the found period gives 49\%, which is the absolute maximum among the mCP candidates in our sample.

\begin{figure}[t]
\centering \resizebox{0.85\hsize}{!}{\includegraphics{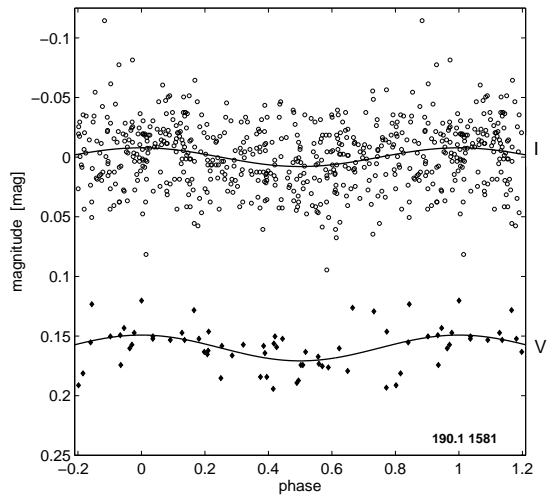}}
\caption{The $V$ and $I$ light curves of star 12 (190.1\,1581) plotted according to the ephemeris given in Table \ref{hodnoty}.} \label{121.24}
\end{figure}

\begin{table}
\begin{center}
\caption{The derived parameters of our light variation model given in Eq. \ref{var}.}\label{hodnoty}
\begin{tabular}{ccrr}
\hline
   & &LMC 190.1 1581 &LMC 190.1 15527 \\
\hline
 \multicolumn{2}{l}{$M_0-2\,450\,000$} &3710.895(36)   & 3710.245(12) \\
 $P$             & [d]      &1.23836(5)             & 0.493570(8) \\
 $\overline{m}_V$& [mag]    &17.8917(16)            & 17.1698(10) \\
 $a_V$           & [mag]    &0.022(4)               & 0.017(3)   \\
$\overline{m}_I$ & [mag]    &17.8814(13)            & 17.1508(8)   \\
 $a_I$           & [mag]    &0.016(4)               & 0.008(2)  \\
\hline
\end{tabular}
\end{center}
\end{table}

\begin{figure}[t]
\centering \resizebox{1\hsize}{!}{\includegraphics{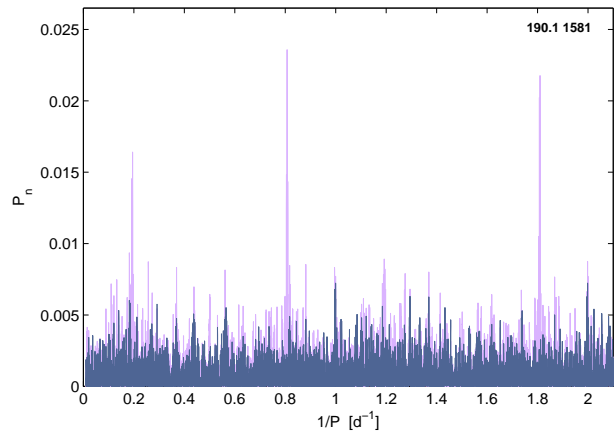}}
\caption{Comparison of the classical Lomb-Scargle periodogram of star 12 (190.1\,1581)  before (pale magenta) and
after subtraction (dark blue) of the variations described by the model (Eq. \ref{var}). It seems that a substantial
part of the variability of the star is well described by this model and its parameters (Table \ref{hodnoty}).} \label{difper12}
\end{figure}

A careful inspection of CCD images of this star reveals that it is a visual binary with a
companion that is about 1.5 magnitude fainter than our target. The OGLE-III photometry was apparently done for both components.
Future photometric observations and treatments of them should take this into account.
On the other hand, the modest ratio of amplitudes of variations in $V$ and $I$ filters completely fulfills
our theoretical expectation.

\subsection{LMC 190.1 15527}

Star 14 (190.1 15527) is the brightest and one of the hottest mCP candidates in our sample. With
$M_{V0}=-1.57$\,mag and $(V-I)_0=-0.09$\,mag it is a late Bp star with Si or He-weak chemical peculiarity. The most
striking result of its analysis is the found period of $P=0\fd494$, which is rather short. If it is
confirmed, it would be one of the fastest rotating mCP stars detected so far.

\begin{figure}[t]
\centering \resizebox{0.85\hsize}{!}{\includegraphics{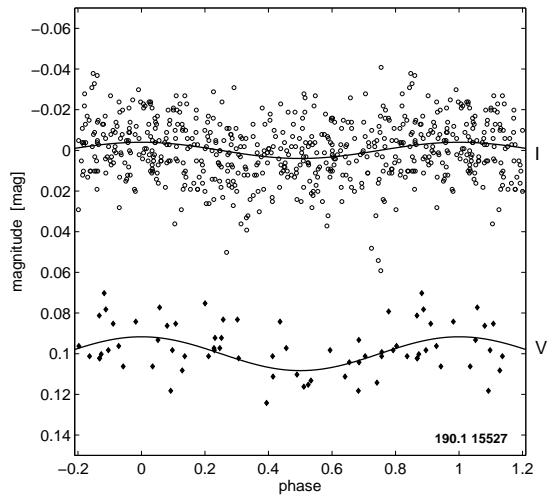}}
\caption{The $V$ and $I$ light curves of star 14 (190.1\,15527) plotted according to the ephemeris given in Table \ref{hodnoty}.} \label{140.49}
\end{figure}

\begin{figure}[t]
\centering \resizebox{1\hsize}{!}{\includegraphics{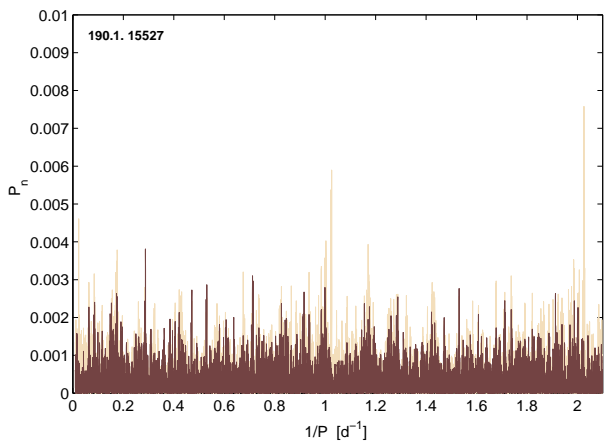}}
\caption{Comparison of the classical Lomb-Scargle periodogram of the star No 14 -- 190.1\,15527 before (pale colour) and after
(dark colour) subtraction of the variations described by the model (Eq. \ref{var}). It seems that a part of the variability
of the star can be described by this model and found parameters (see Table \ref{hodnoty}). Unfortunately, there are some indications
that the main peak may be an artefact caused by specifically distributed data.} \label{difper14}
\end{figure}

Unfortunately, the found period shorter than 12 hours is not as firmly established as in the case of the
previous mCP candidate with larger light changes. It results in the relatively low appreciation of the reality probability of the
found period only 30\%. The rather small variations in the $I$ filter (see Fig. \ref{140.49}) are
also not in favour of variability, nor are our simulations with randomly distributed magnitudes that also constitute periodogram
peaks slightly larger than 2 and 1. That is why we suspect the feature to be an artefact caused by the specifically distributed
observations. However, Fig. \ref{difper14} shows the success of the expression of the light curves by the simple model given in Eq.\,\ref{var}.

In conclusion, we suggest that the periodic variations of star 14 (190.1 15527) remains an open question.

\section{Conclusions}

We analysed the OGLE-III photometry for fourteen stars of which
twelve are mCP candidates. In total, 6476 individual photometric measurements
in $V$ and $I$ were used to perform a time series analysis. For this purpose,
we calculated periodograms including a bootstrap-based probability of the
statistical significance.
Although the median of the amplitudes of maximum peaks, $\mathrm{median}\,
A_{\rm{m}}=0.02$\,mag in the periodograms of the mCP candidates, does not
contradict our expectation \citep[$\mathrm{median} (A_{\rm{m}})=0.032$\,mag
for Galactic mCP stars;][]{zoo}, other circumstances suggest that the prevailing
majority of these peaks might not be statistically significant.
We found some periodic variations in only two of the mCP candidates, of which
one is doubtful.

It seems that the rotationally modulated variability of the studied
mCP candidates in the $V$ and $I$ bands is very weak (if present at all).
As upper limits of their effective amplitudes, we used $A_{\rm{effs}}$ derived from
the analysis of shuffled data. Nevertheless, it is very likely that the
true amplitudes are much lower than 0.01 mag, which is the limit derived from the
time series analysis of the brightest mCP candidate.

From this finding we can conclude that the spots on the stellar
surfaces of these objects are not as contrasting as those of their
Galactic counterparts; in other words, the overabundant elements
(Si, Fe, Cr, and so on) are more evenly distributed over the
surface or the maximum abundance is lower. This phenomenon has also been
observed for non-magnetic Galactic Am or weakly magnetic HgMn stars.

The low contrast of LMC mCP photometric spots could be
explained by lower global magnetic fields strengths of LMC mCP stars
than their Galactic counterparts.
It may relate with the lower level of interstellar magnetic field in
the LMC if we compare it with the Galactic field, which results
in the overall lower percentage of mCP star appearance compared
to the Milky Way \citep{pau06}.

Our results could also have a fundamental impact on the
origin of the global stellar magnetic fields for those objects.
Two theories have been developed for this \citep{mos89}, and are still in dispute. The fossil theory has two variants:
the magnetic field is either the slowly decaying relic of the
frozen-in interstellar magnetic field or of the dynamo acting
in the pre-main sequence phase. The dynamo theory is based on the existence
of a contemporaneous dynamo operating in the convective core of the
magnetic stars. If we assume that the stellar dynamo functions with the
same efficiency for all stars with an identical mass and luminosity,
than we should find the same number of mCP stars in the Milky Way
and the LMC. On the other hand, the global magnetic field of the
LMC \citep[$\approx$1.1\,$\mu$G,][]{gae05} is much weaker than that of the
Milky Way \citep[6 -- 10\,$\mu$G,][]{beck09}.
Therefore, our results are in favour of the fossil
theory being the cause of the CP star phenomenon.

Another important global parameter which has to be taken into
account is the metallicity. The overall metallicity of the LMC
is about $-$0.5 to $-$2.0\,dex lower than that of the Milky Way.
For the CP phenomenon, in general, besides Fe, Mg, Si, and Cr are
the most important elements for the light variability and thus
the spot characteristics. Our targets are members of or in the
surrounding of the open clusters NGC 1711, NGC 1866, and NGC 2136/7.
In addition, one field in the bulge was observed. The published
[Fe/H] values for those regions are all below $-$0.70\,dex
\citep{pau06}. \citet{pom08} published detailed [$\alpha$/Fe]
ratios for the inner disc of LMC. They conclude that the
abundances of Mg, Si, Ti, and Cr scale the same way as [Fe/H].
The intrinsic abundance of our targets should be, therefore, uniform,
not favouring any element for magnetic diffusion and thus the spot
characteristics or the maximum elemental abundances are too low to cause
any substantial light variability.

As future steps, we suggest retrieving time series of the remaining
published mCP candidate stars and getting more accurate measurements
of the presented stars. In addition, observations in other wavelength
regions would shed more light on the surface characteristics.

\begin{acknowledgements}
This work was supported by the following grants: GA \v{C}R
P209/12/0217, 7AMB12AT003, WTZ CZ-10/2012, \#LG12001 (Czech Ministry
of Education, Youth and Sports), and FWF P22691-N16.
The OGLE project has received funding from the European
Research Council under the European Community's Seventh Framework
Programme (FP7/2007-2013)/ERC grant agreement No. 246678. We dedicate
this paper to Willy Schreiner who tragically died during its writing.
\end{acknowledgements}

\Online
\begin{figure}[t]
\centering \resizebox{0.85\hsize}{!}{\includegraphics{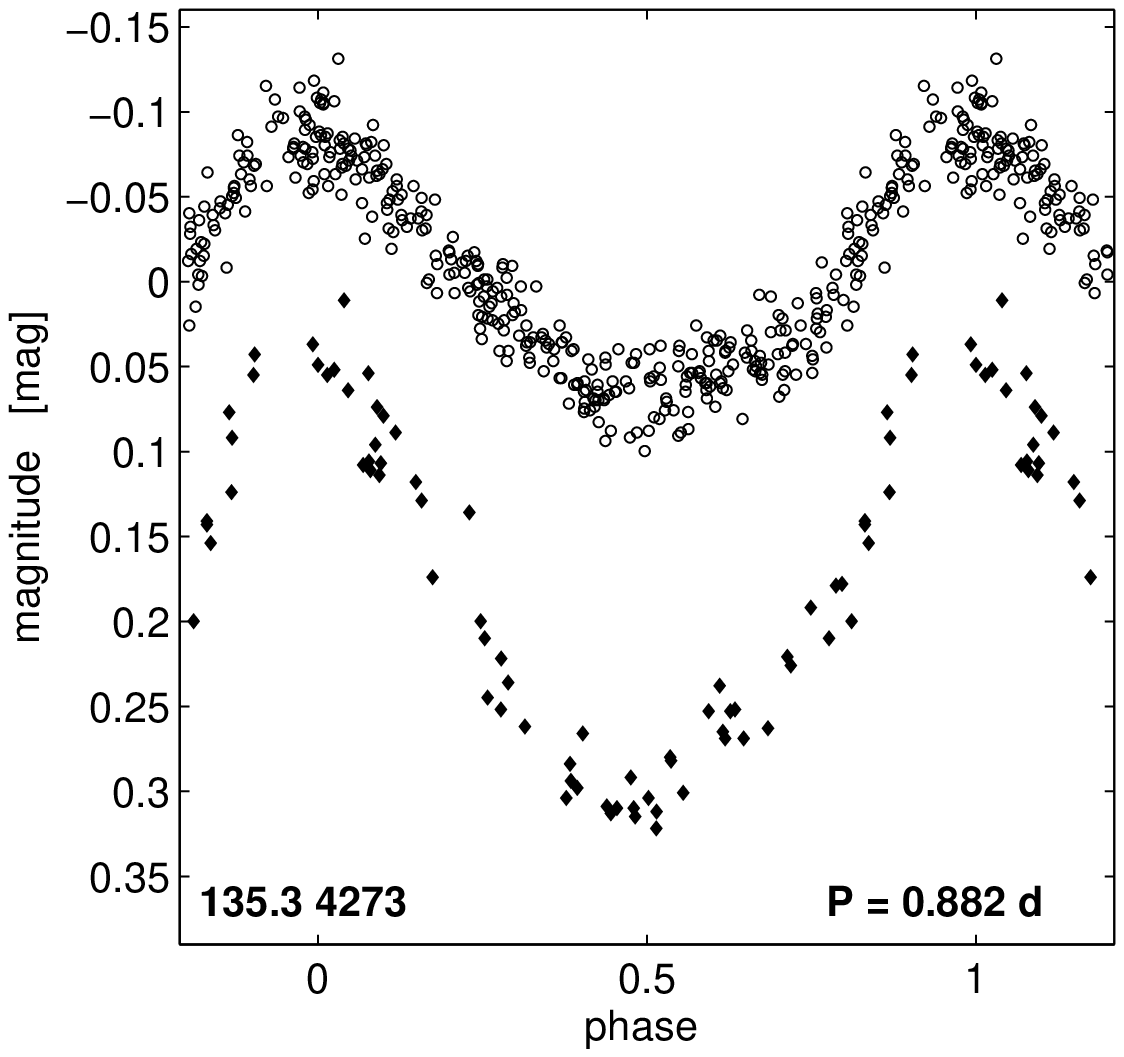}}
\caption{The $V$ ($\blacklozenge$) and $I$ ($\circ$) light curves of star 1 (135.3\,4273) plotted according to the ephemeris given in Table \ref{hodnoty}.}
\end{figure}
\begin{figure}[t]
\centering \resizebox{0.85\hsize}{!}{\includegraphics{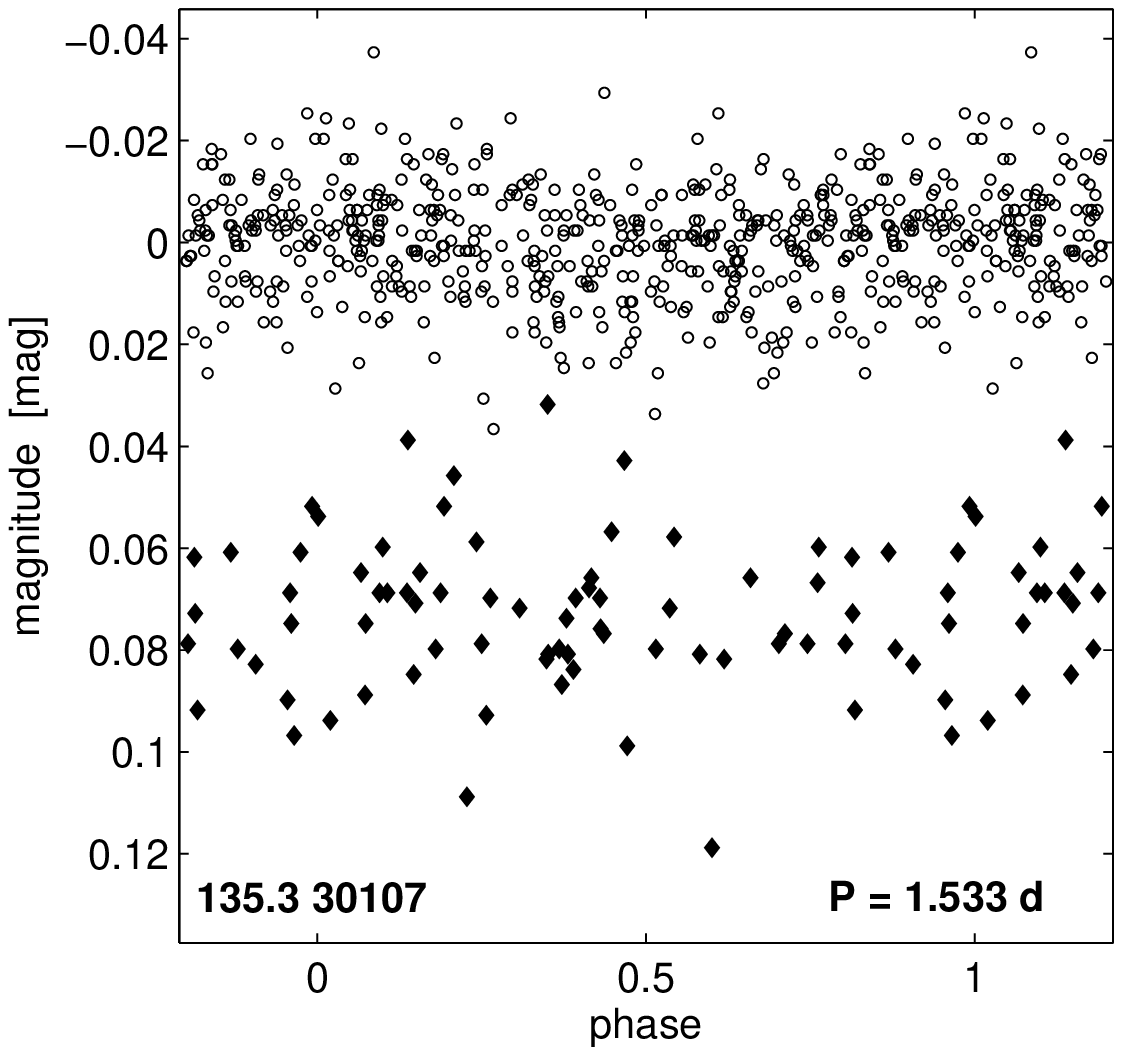}}
\caption{The $V$ ($\blacklozenge$) and $I$ ($\circ$) light curves of star 2 (135.3\,30107) plotted according to the ephemeris given in Table \ref{hodnoty}.}
\end{figure}
\begin{figure}[t]
\centering \resizebox{0.85\hsize}{!}{\includegraphics{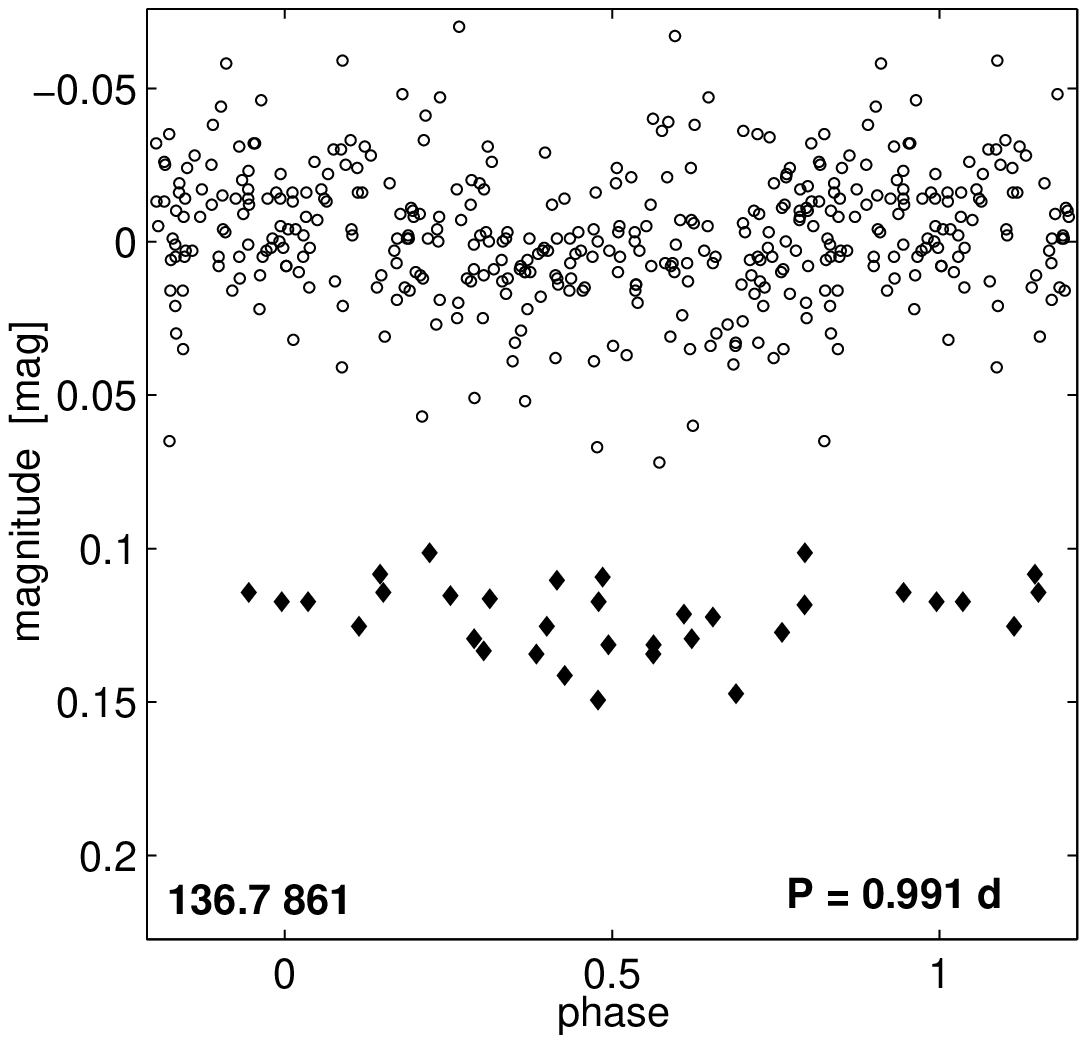}}
\caption{The $V$ ($\blacklozenge$) and $I$ ($\circ$) light curves of star 3 (136.7\,861) plotted according to the ephemeris given in Table \ref{hodnoty}.}
\end{figure}
\begin{figure}[t]
\centering \resizebox{0.85\hsize}{!}{\includegraphics{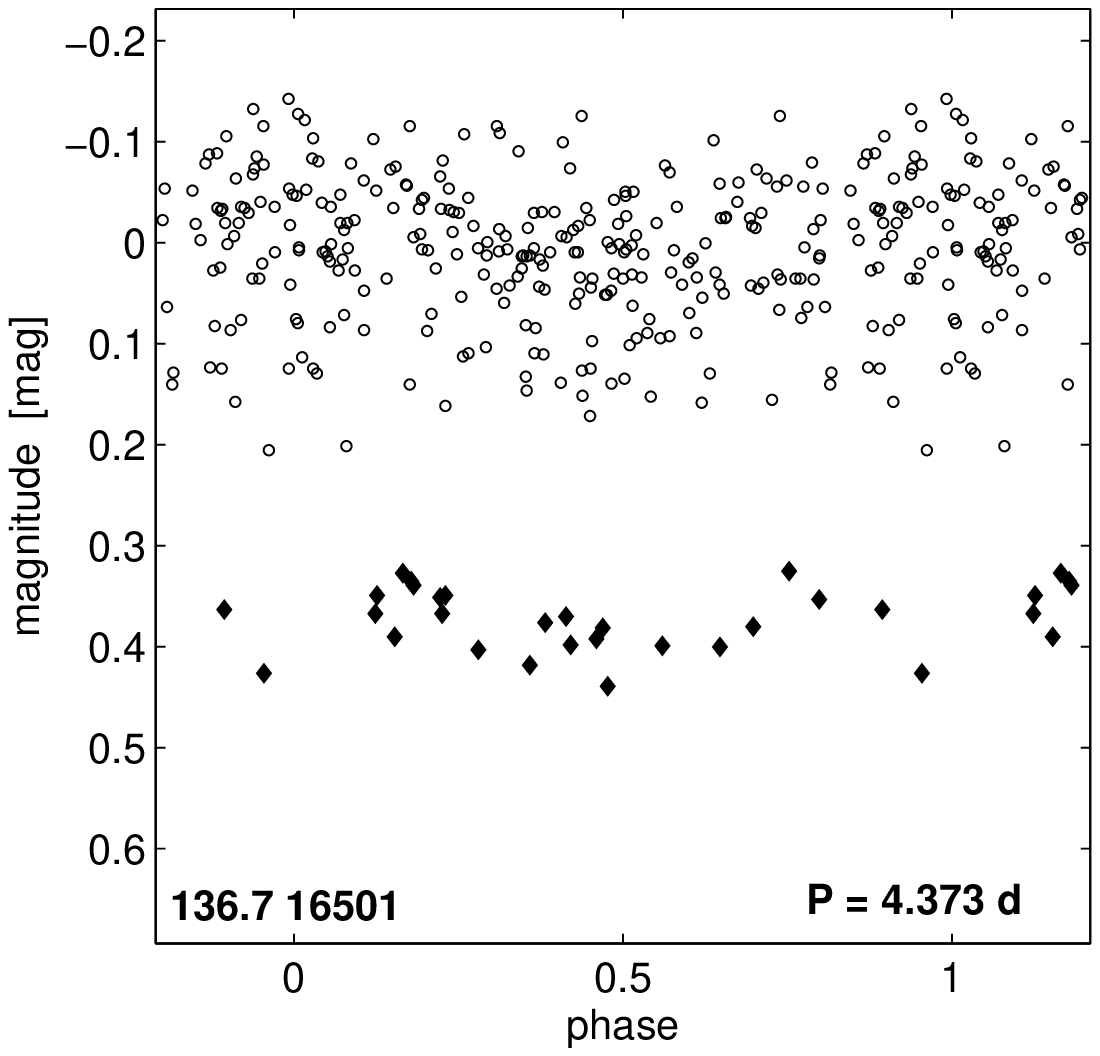}}
\caption{The $V$ ($\blacklozenge$) and $I$ ($\circ$) light curves of star 4 (136.7\,16501) plotted according to the ephemeris given in Table \ref{hodnoty}.}
\end{figure}
\begin{figure}[t]
\centering \resizebox{0.85\hsize}{!}{\includegraphics{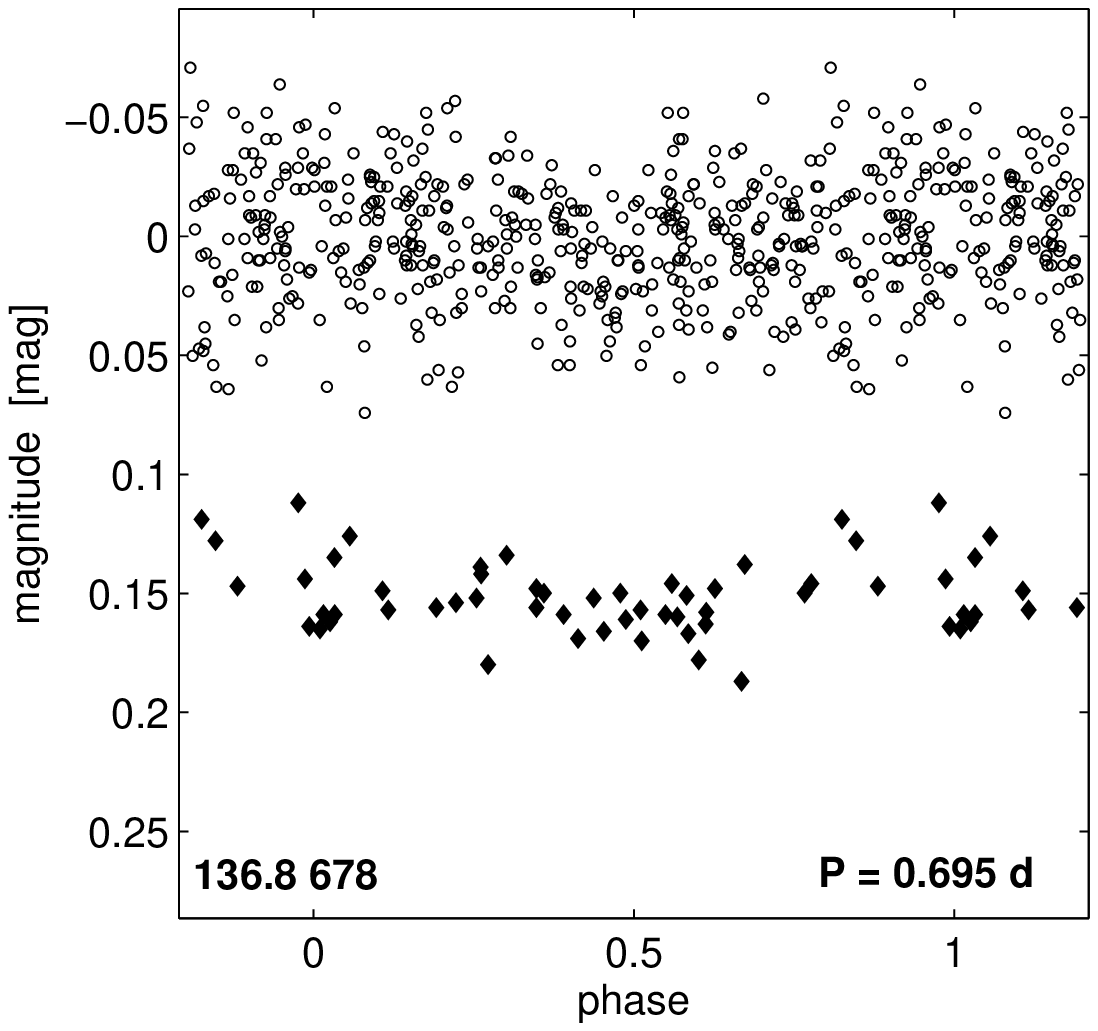}}
\caption{The $V$ ($\blacklozenge$) and $I$ ($\circ$) light curves of star 5 (136.8\,678) plotted according to the ephemeris given in Table \ref{hodnoty}.}
\end{figure}
\begin{figure}[t]
\centering \resizebox{0.85\hsize}{!}{\includegraphics{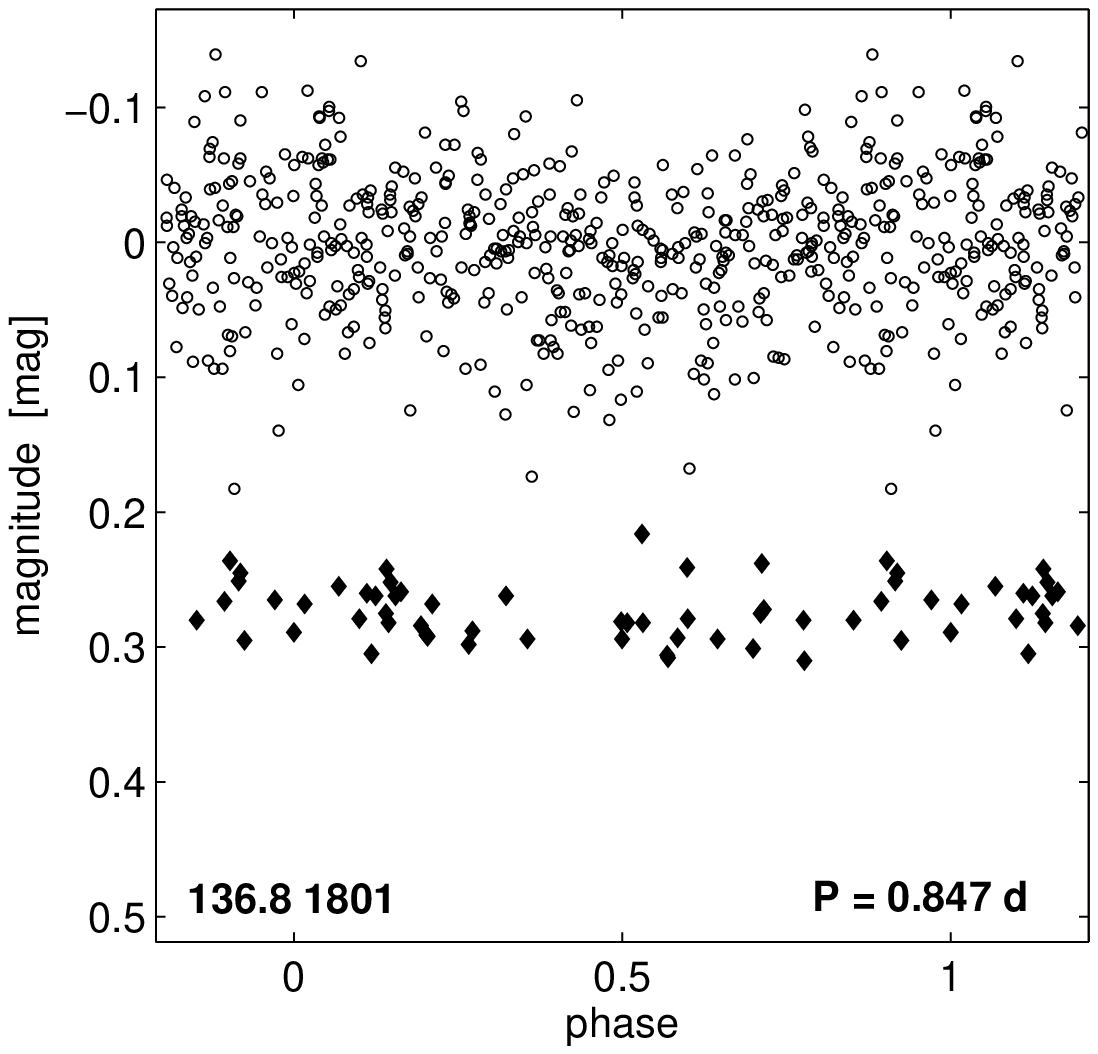}}
\caption{The $V$ ($\blacklozenge$) and $I$ ($\circ$) light curves of star 6 (136.8\,1801) plotted according to the ephemeris given in Table \ref{hodnoty}.}
\end{figure}
\begin{figure}[t]
\centering \resizebox{0.85\hsize}{!}{\includegraphics{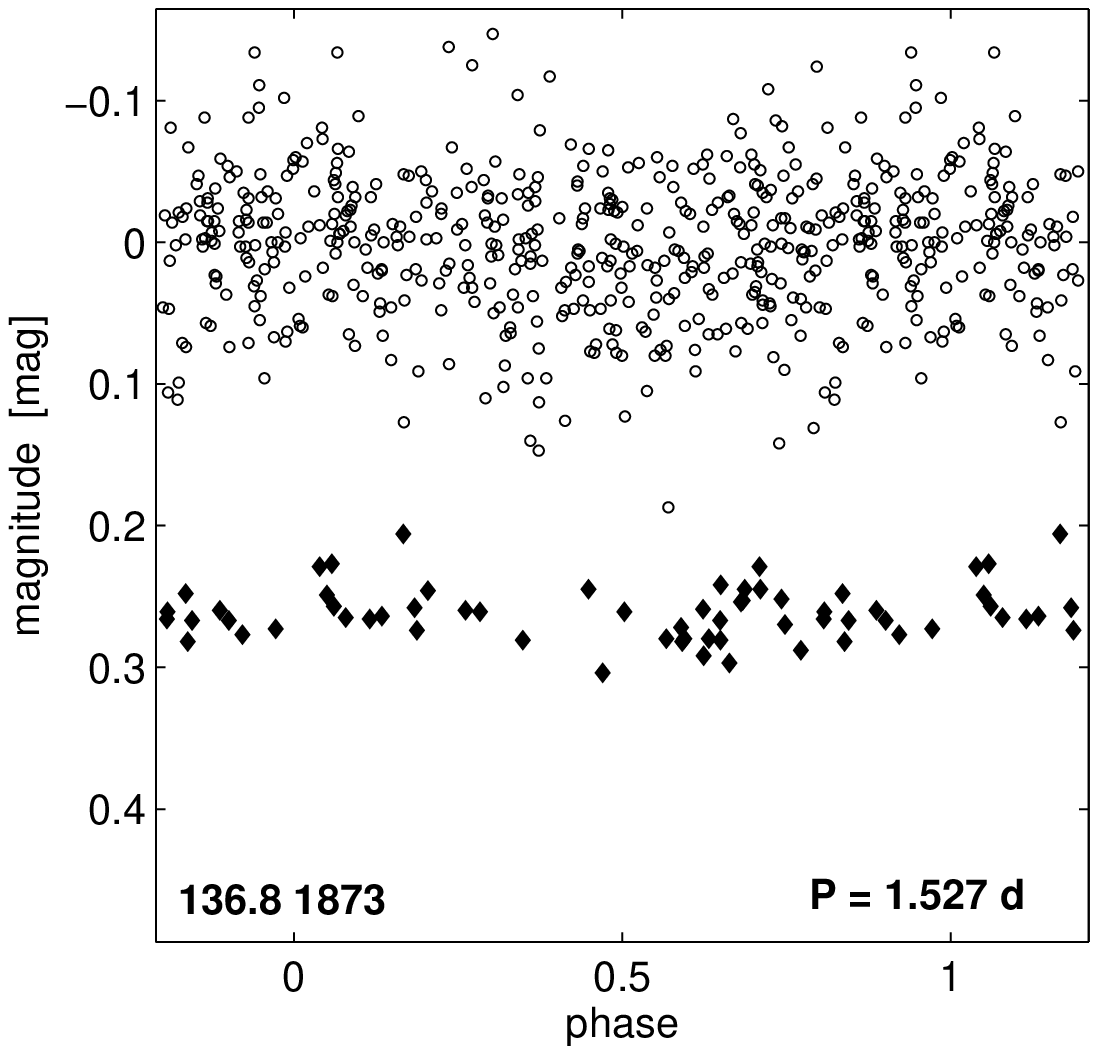}}
\caption{The $V$ ($\blacklozenge$) and $I$ ($\circ$) light curves of star 7 (136.8\,1873) plotted according to the ephemeris given in Table \ref{hodnoty}.}
\end{figure}
\begin{figure}[t]
\centering \resizebox{0.85\hsize}{!}{\includegraphics{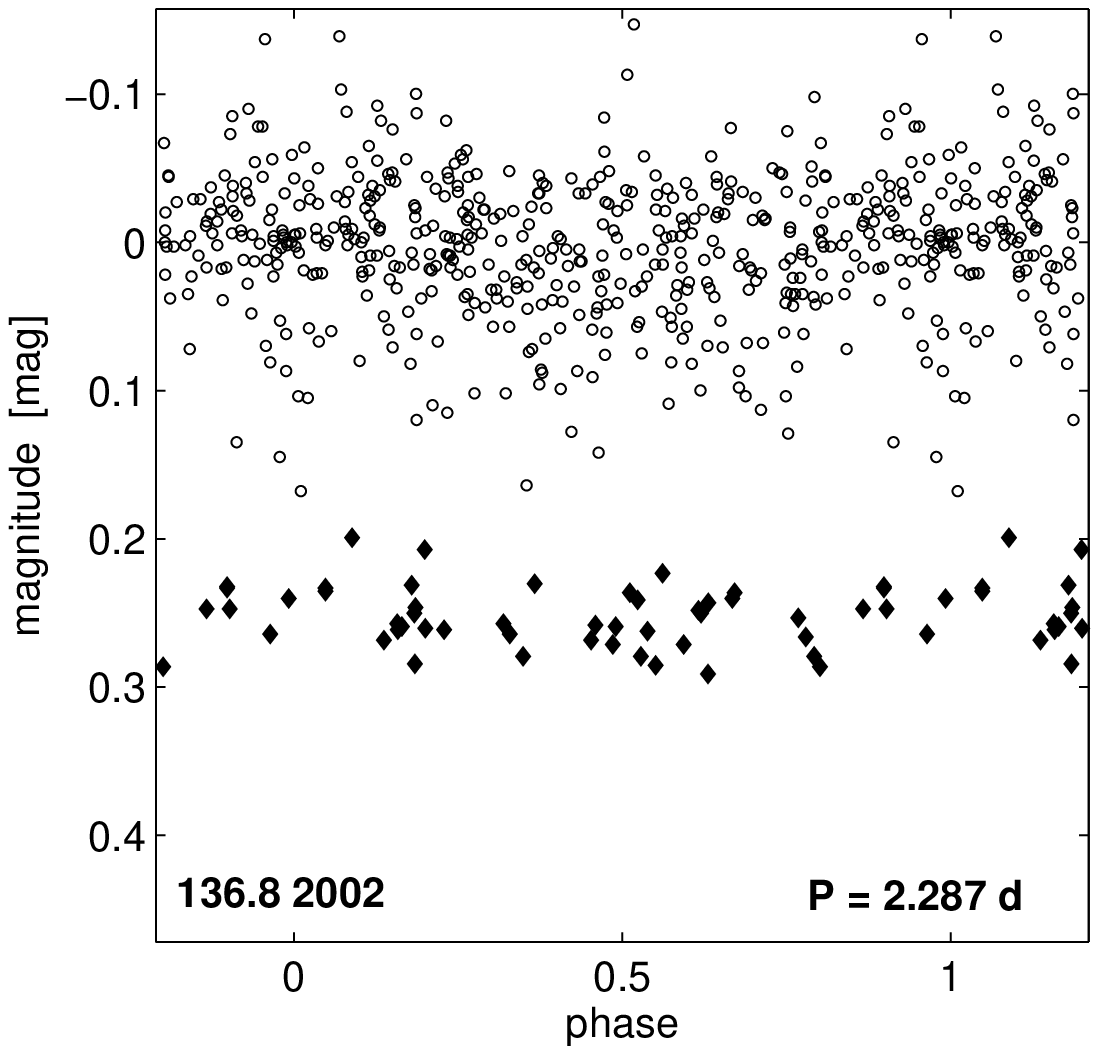}}
\caption{The $V$ ($\blacklozenge$) and $I$ ($\circ$) light curves of star 8 (136.8\,2002) plotted according to the ephemeris given in Table \ref{hodnoty}.}
\end{figure}
\begin{figure}[t]
\centering \resizebox{0.85\hsize}{!}{\includegraphics{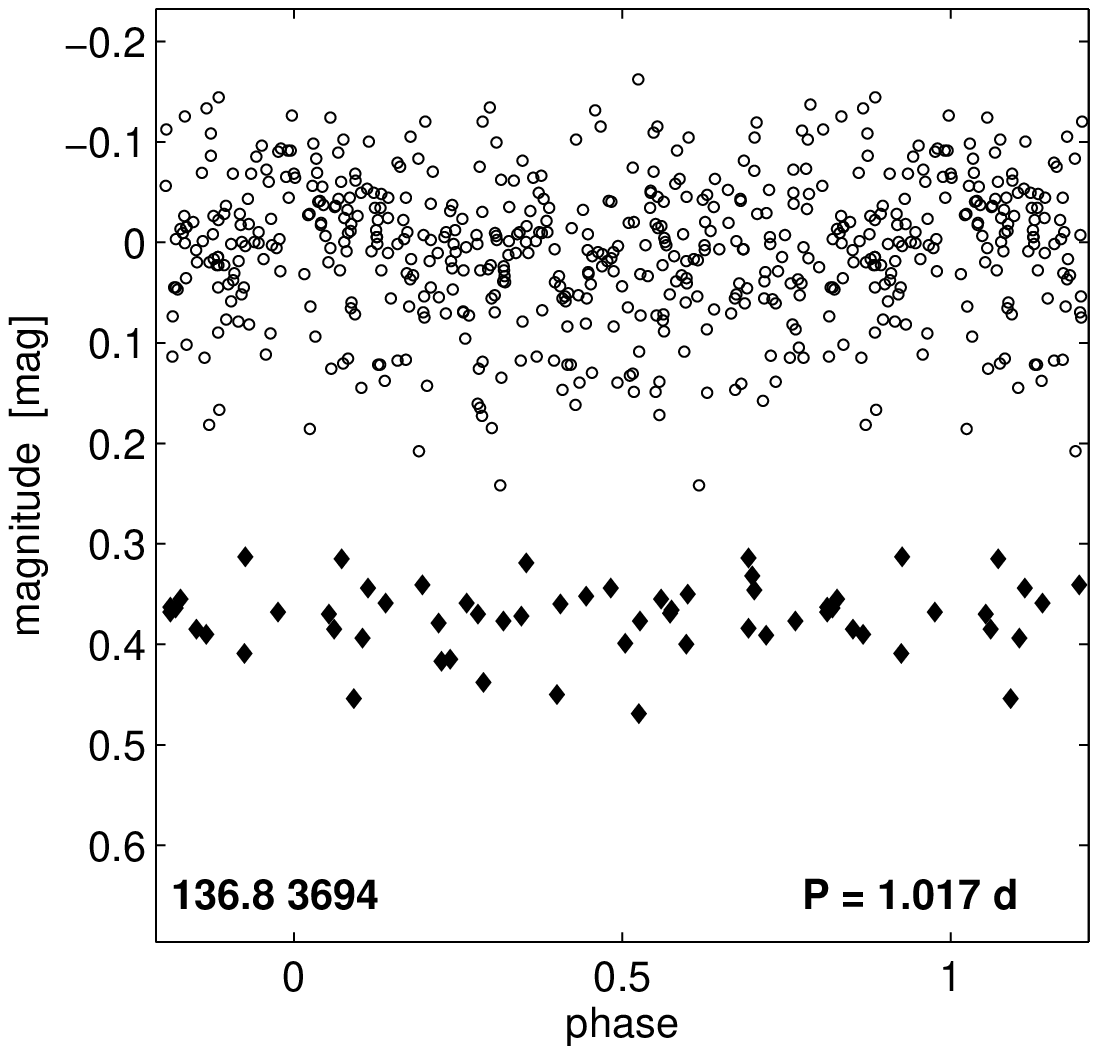}}
\caption{The $V$ ($\blacklozenge$) and $I$ ($\circ$) light curves of star 9 (136.8\,3694) plotted according to the ephemeris given in Table \ref{hodnoty}.}
\end{figure}
\begin{figure}[t]
\centering \resizebox{0.85\hsize}{!}{\includegraphics{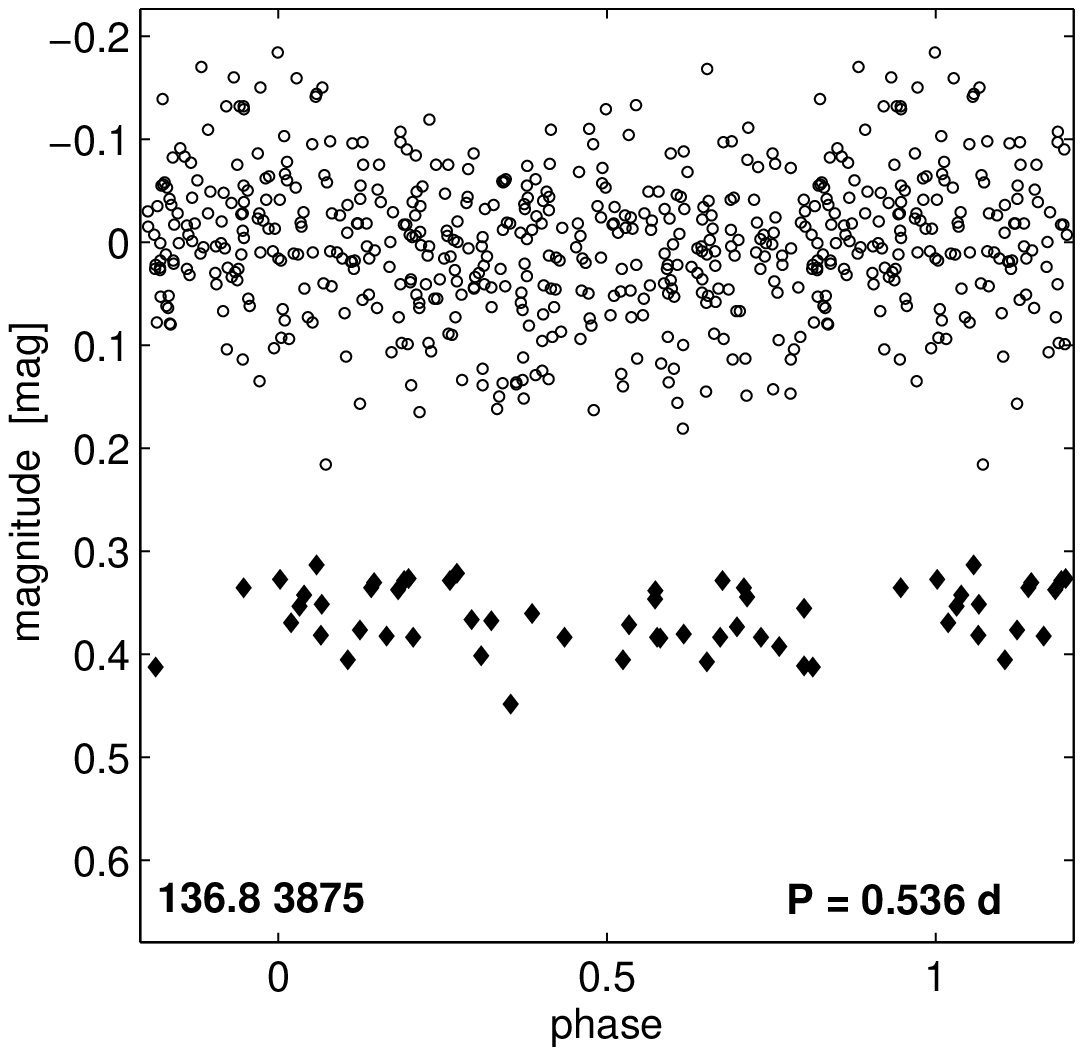}}
\caption{The $V$ ($\blacklozenge$) and $I$ ($\circ$) light curves of star 10 (136.8\,3875) plotted according to the ephemeris given in Table \ref{hodnoty}.}
\end{figure}
\begin{figure}[t]
\centering \resizebox{0.85\hsize}{!}{\includegraphics{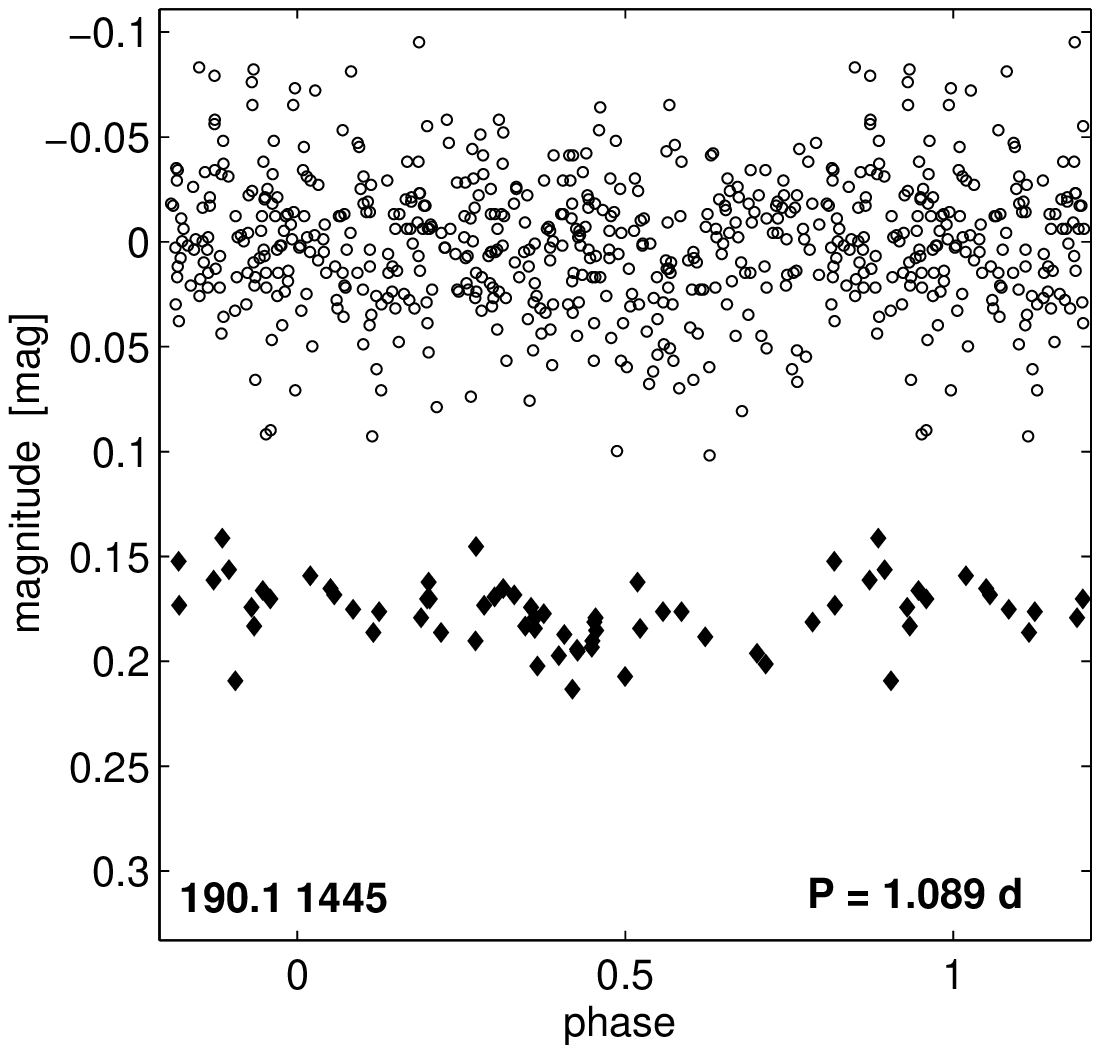}}
\caption{The $V$ ($\blacklozenge$) and $I$ ($\circ$) light curves of star 11 (190.1\,1445) plotted according to the ephemeris given in Table \ref{hodnoty}.}
\end{figure}
\begin{figure}[t]
\centering \resizebox{0.85\hsize}{!}{\includegraphics{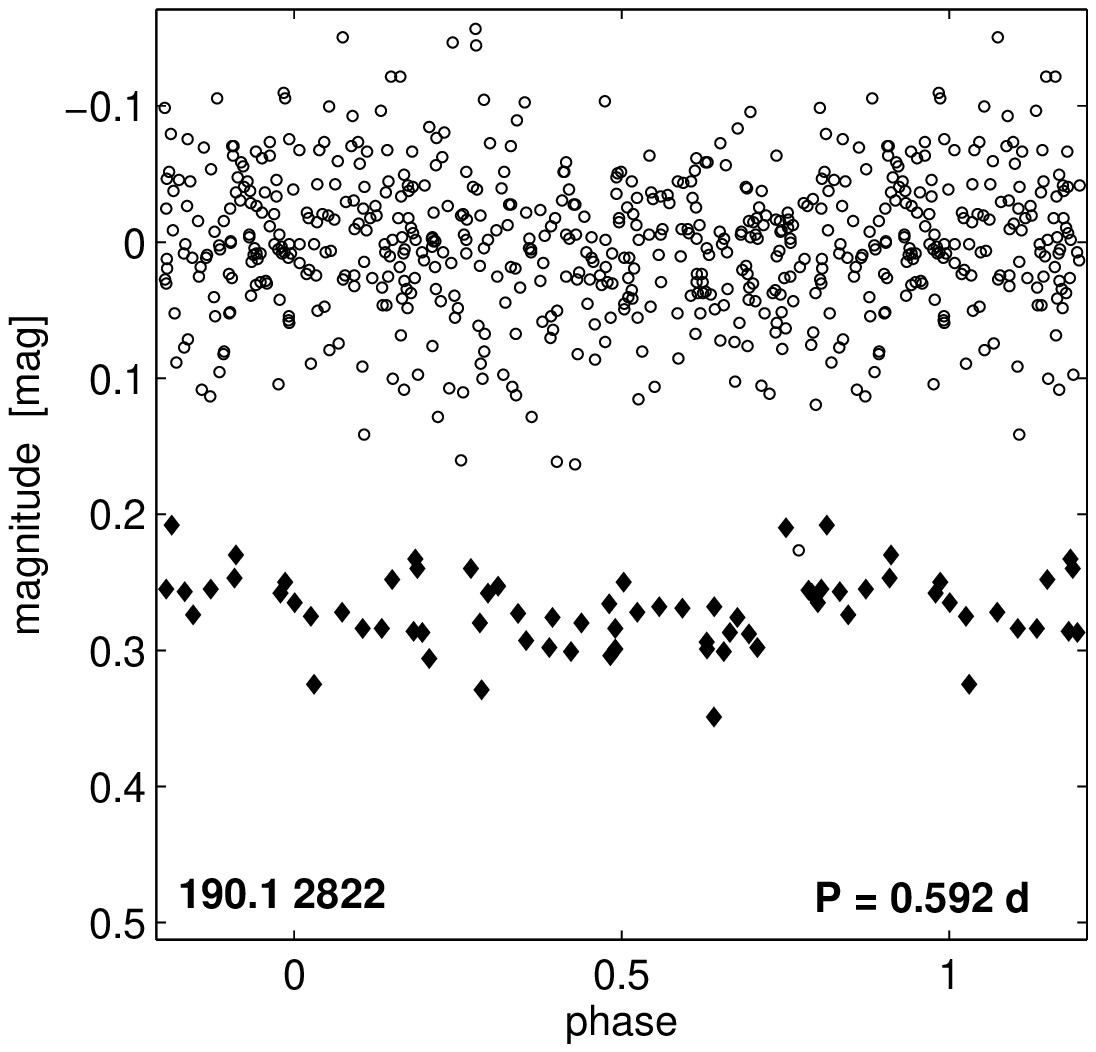}}
\caption{The $V$ ($\blacklozenge$) and $I$ ($\circ$) light curves of star 13 (190.1\,2822) plotted according to the ephemeris given in Table \ref{hodnoty}.}
\end{figure}
\end{document}